\documentclass{iaus}
\usepackage{graphicx}
\def\hhco       {H$_2$CO}

\def\decdeg {\rlap . {}^\circ}     %e.g. $40\decdeg 5$ for 40.5 degrees
\def\JK         {$J_{K_{\rm a},K_{\rm c}}=$}
\def\Jk         {$J_k=$}
\def\pJk        {\phantom{$J_k=$}}

\def\JKN        {$(J,K)=$}
\def\pJKN       {\phantom{$(J,K)=$}}

\def\d {\phantom{$0$}}

\def\hzo        {H$_2$O}
\def\meth   {CH$_3$OH}
\def\nhhh       {NH$_3$}
\def\deg      {\ifmmode{^{\circ}}\else{$^{\circ}$}\fi}
\def\as     {\ifmmode {\rlap.}$\,$''$\,$\! \else ${\rlap.}$\,$''$\,$\!$\fi}
\def\am     {\ifmmode {\rlap.}$\,$'$\,$\! \else ${\rlap.}$\,$'$\,$\!$\fi}
\def\kms    {\ifmmode{{\rm km~s}^{-1}}\else{km~s$^{-1}$}\fi}

\title{Exciting Maser Science with New Instruments %in the near and the far future --
-- the Promise of the EVLA}

\author{Karl M. Menten}
\affiliation{Max-Planck-Institut f\"ur Radioastronomie, Auf dem H\"ugel 69, D-53121, Germany}

\pubyear{2007}
\volume{242}  %% insert here IAU Symposium No.
\pagerange{1--10}
\date{?? and in revised form ??}
\jname{Astrophysical masers and their environments }
\editors{J. Chapman \& W. Baan, eds.}
\begin{document}

\maketitle

\begin{abstract}
In the near future, the Expanded Very Large Array (EVLA) will allow
surveys for maser
sources with unprecedented sensitivity, spectral coverage and spectroscopic capabilities.
In particular, comprehensive surveys for many maser species with
simultaneous sensitive continuum imaging and absorption studies will
give a comprehensive radio picture of star formation in the Galactic plane
and elsewhere.
Very efficient EVLA surveys for H$_2$O megamasers in Active Galacic Nuclei will be possible to practically arbitrary redshifts.

EVLA and Atacama Large Millimeter Array (ALMA) studies of H$_2$O and SiO masers will
serve as high resolution probes of the innermost envelopes of
oxygen-rich evolved stars and HCN masers of carbon-rich stars.

Farther in the future, the Square Kilometer Array (SKA) promises
the detection of OH gigamasers at all conceivable redshifts and maser astrometry with unprecedented accuracy.

\keywords{masers, instrumentation: high angular resolution,
instrumentation: interferometers, surveys, stars: formation,
Galaxy: structure
}
\end{abstract}

\firstsection % if your document starts with a section,
\section{\label{intro}Introduction -- Maser Surveys}

Arguably, one of the main uses for astronomical masers is their signpost function. To give a few examples: Interstellar water vapor (\hzo) masers arise from high velocity outflows of accreting low- and high-mass protostars. Many class II methanol (\meth) masers mark the positions of deeply embedded protostars that are difficult or frequently impossible to locate by any other phenomenon. 1612 MHz hydroxyl (OH) masers mark intermediate-mass stars during a short-lived phase of high mass loss.
Surveys for maser emission thus help finding interesting sources and often have surprising results.
%They can be either unbiased, \textit{``blind''} surveys covering promising areas of the sky, e.g., molecular %clouds or parts or all of the Galactic plane or \textit{targeted} surveys of objects selected by certain %criteria.

\subsection{Two Types of Surveys}
There are, essentially, two types of surveys - \textit{Blind surveys} and
\textit{targeted surveys}.
\subsubsection{\label{blindsurveys}Blind Surveys}
\noindent
Blind surveys cover large areas of the sky in an unbiased way, integrating (possibly) uniformly to a pre-set flux density limit. The covered area may be the whole sky, like for the NRAO Very Large Array Sky Survey (NVSS), a 1.4 GHz continuum survey of all the sky north of $-40\deg$ declination ({\cite{Condon_etal1998}), but for maser surveys they usually are not. Rather, most blind maser surveys cover selected molecular clouds or portions of the Galactic plane.
An important example of the latter is the unbiased OH maser survey of the southern Galactic plane conducted by Caswell and collaborators %(\cite{Caswell_etal1980, CaswellHaynes1983, Caswell_etal1987}).
(\cite{Caswell_etal1987} and references therein).

Surveys for \hzo\ masers in the spiral arms of external galaxies may cover whole galaxies, which for nearby systems such as M31 or M33 comprise many hundreds or even thousands of separate pointings. Blind surveys can be made either with multiple separate pointings, measuring, for full sampling, a grid of $0.5\times$FWHM beam-width spaced positions. %\footnote{The full width at half maximum (FWHM) size of a telescope's main beam is given by %$\approx1.22\lambda/D$, where $\lambda$ is the observing frequency and $D$ is the diameter of the %telescope's aperture.}.
Alternately and more efficiently, ``on-the-fly'' mapping may be employed, during which the telescope is constantly moved at a certain slew rate and data are taken continuously.
In the past, most blind surveys were conducted using single dish telescopes. For newly found sources, in many cases improved positions were determined with follow-up observations with accuracies limited by telescope pointing errors (usually a few arcseconds) or much smaller ($\ll 1''$) with interferometers.

Clearly, full-interferometric surveys are highly preferable (see \S\ref{vla}), delivering much better gain stability, position accuracy and direct information on the small-scale maser distributions. However, the limited correlator bandwidth of today's most effective interferometer, the NRAO Very Large Array (VLA), has been a severe problem in conducting efficient blind surveys at high frequencies, in particular for sources with a priori unknown velocities (see \S\ref{frequencycoverage}).

\subsubsection{\label{targeted}Targeted Surveys}
In contrast, \textit{targeted surveys} consist of pointed observations toward sources that were identified as promising targets based on certain criteria.
If one searches for masers in a certain species, one such criterion might be the existence of maser emission from another species at or in the vicinity of the target position. Examples for this are searches toward the positions of the OH masers found in the Caswell surveys (\S\ref{blindsurveys}), for  \hzo\
masers %(\cite{Caswell_etal1983a, Caswell_etal1983b, Caswell_etal1989})
(\cite{Caswell_etal1989} and references therein)
and \meth\ masers (\cite{Norris_etal1987, Menten1991, Caswell_etal1995}).

Another example is extragalactic \hzo\ masers, all of which were either found in targeted searches toward HII regions complexes (first in M33; \cite{Churchwell_etal1977}) or the nuclear regions of the target galaxies. The latter masers are located either in a parsec-scale accretion disk around an active galactic nucleus  or interacting with a jet emerging from it (see, e.g., \cite{Lo2005} and Greenhill, these proceedings) or from a starburst region surrounding the nucleus on a few hundred pc scale (e.g., Castangia et al., these proceedings).

Extremely fruitful selection criteria for finding masers and the interesting sources that host them have been color criteria derived from mid/far-infrared surveys. Namely, the InfraRed Astronomy Satellite (IRAS) covered the whole sky at 12, 25, 60, and 100$\mu$m, which resulted in the IRAS Point Source Catalogue. A survey of a $\pm 5\deg$ latitude wide strip of the whole Galactic plane was made with the Midcourse Satellite Experiment, MSX, in several IR bands. %(\cite{Price1995}).
Representative portions of the plane were covered by other surveys, e.g., ``ISOGAL'' with the Infrared Space Observatory, ISO (\cite{Omont_etal2003, Schuller_etal2003}). New multi-IR band space-borne surveys of the Galactic plane and the whole sky with the Spitzer Space Observatory (GLIMPSE, \cite{Benjamin_etal2003}) and Akari (\cite{Murakami_etal2007}), respectively, will yield many thousands of targets for maser surveys.
We now give a few examples for IR color-selected maser surveys.

A series of successful 1612 MHz OH satellite line surveys with the Australia Telescope Compact Array (ATCA) and the VLA using IRAS colors resulted in the detection of more than a thousand OH/IR stars %, i.e., evolved stars in the very high-mass loss end of the asymptotic giant branch (AGB) lifetime
%(e.g. %\cite{Sevenster_etal1997a, Sevenster_etal1997b, Sevenster_etal2001}).
(\cite{Sevenster_etal2001} and references therein).

Deguchi and collaborators have highly effectively used IRAS colors to identify targets for very successful 43 GHz $v = 1$ and $2, J = 1 \to 0$ SiO and 22.2 cm \hzo\ maser surveys of practically all of the Galactic plane accessible with the  Nobeyama 45 m telescope.
Using infrared magnitudes and colors from the ISOGAL and MSX catalogues,
\cite{Messineo_etal2002} searched for SiO maser emission in 441 objects and detected it in 271 of them.

While all of the above surveys found many hundreds of objects allowing statistical and kinematical studies (like characterizing the Galactic bar potential; see \cite{Habing_etal2006} and H. Habing's contribution to these proceedings), they also resulted in the serendipitous discovery of a number of highly interesting objects. For example, 1612 MHz OH maser sources with extremely wide and/or peculiar velocity appearance could be classified as extreme mass-loss proto-planetary nebulae: important, short-lived and, therefore, difficult to find transition objects in the evolution from asymptotic giant branch (AGB) star to planetary nebula (e.g.,
\cite{Zijlstra_etal2001, Deacon_etal2007}; see also J. Chapman's contribution to these proceedings).

\begin{figure}
\begin{center}
\includegraphics[width=6cm]{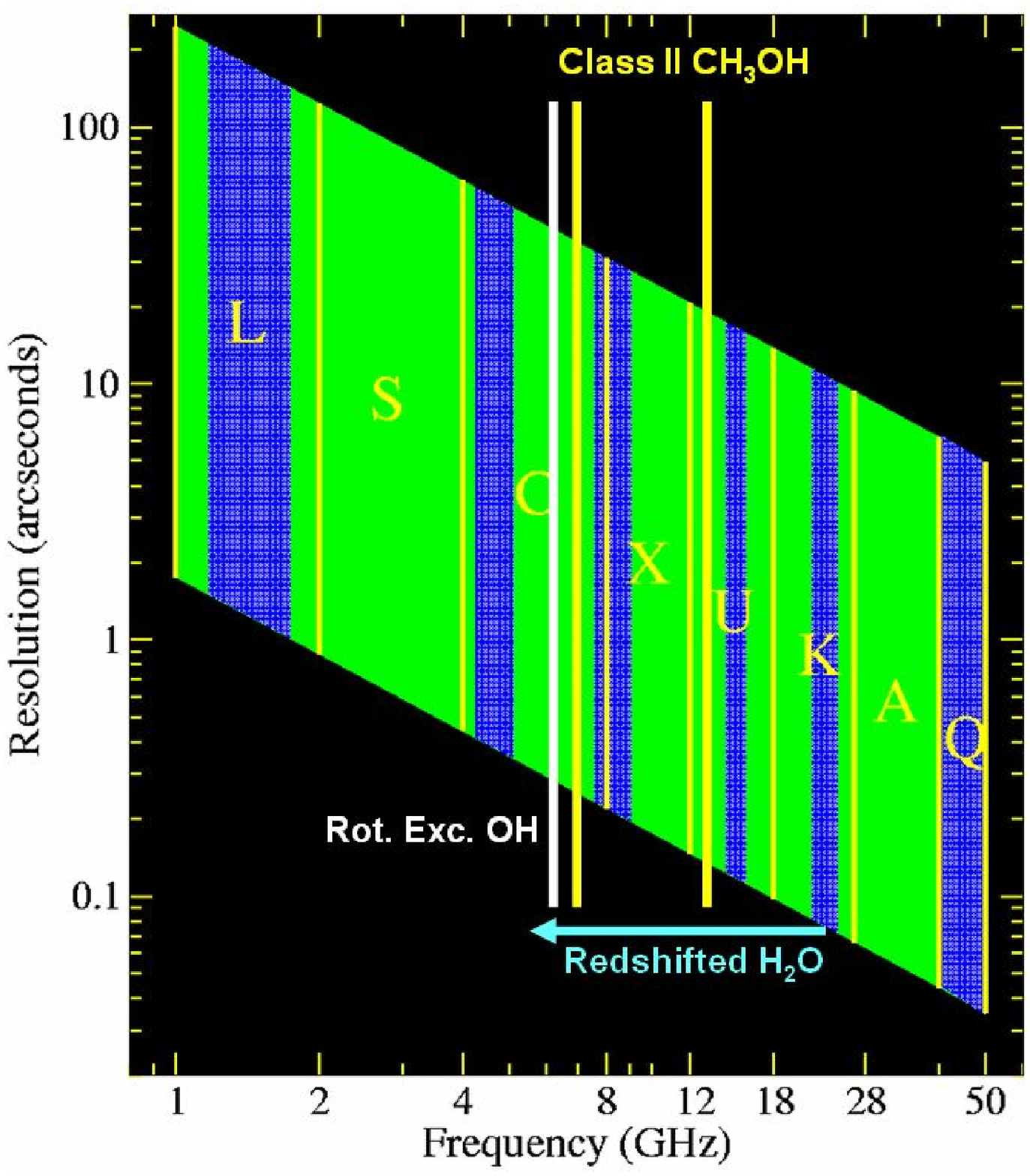}
\includegraphics[width=7cm]{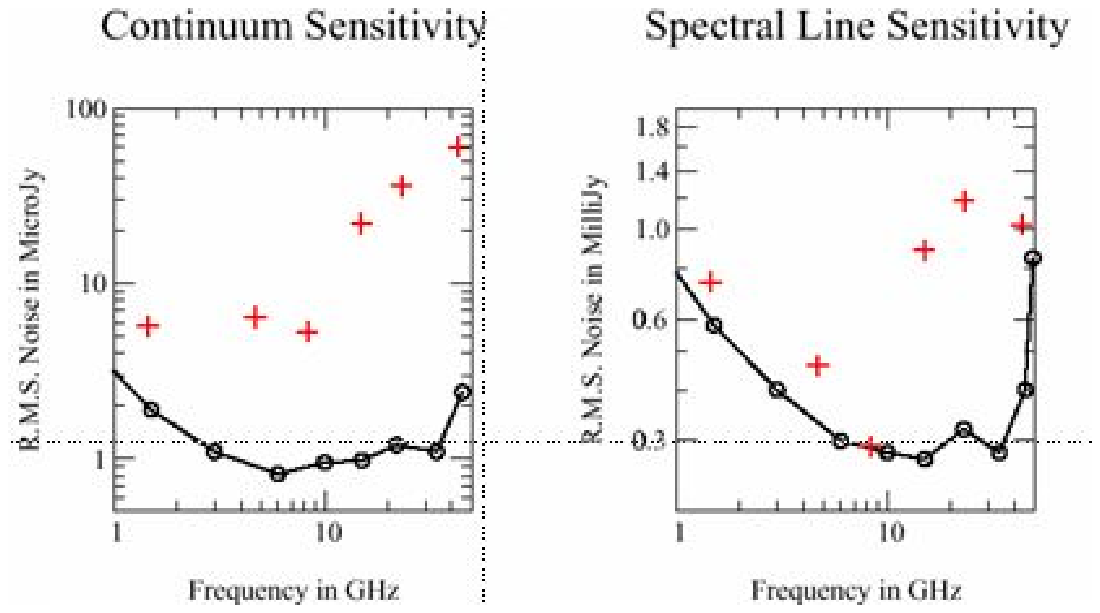}
  \caption{\textit{Left}: Comparison of the frequency coverage of the present VLA (dark stripes) with that of the EVLA (dark plus lighter stripes, i.e. continuous coverage from 1 to 50 GHz). The frequencies of the 6.7 and 12.2 GHz \meth\ and the 6.0 GHz OH lines are indicated. Also indicated is the EVLA's virtually limitless redshift coverage of the 22.2 GHz \hzo\ line. Note that the \hzo\ line from the $z=0.660$ QSO SDS J0804+3607 is redshifted to 13.4 GHz (\cite{BarvainisAntonucci2005}).
  \textit{Right}: Comparison of the sensitivities of the present VLA (crosses) and the EVLA (continuous line). Shown is the $1\sigma$ rms noise level reached in 12 h of integration for continuum and spectral line observations in a 1 \kms\ wide velocity channel. Figures are taken from the EVLA proposal, which is available on the EVLA WWW site: http://www.aoc.nrao.edu/evla/pbook.shtml.} \label{fig:evlacoverage}.
\end{center}
\end{figure}

\begin{figure}
\begin{center}
\includegraphics[width=6.5cm]{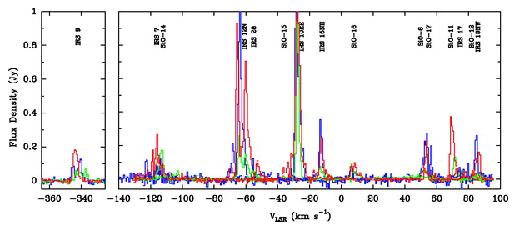}
\includegraphics[width=6.5cm]{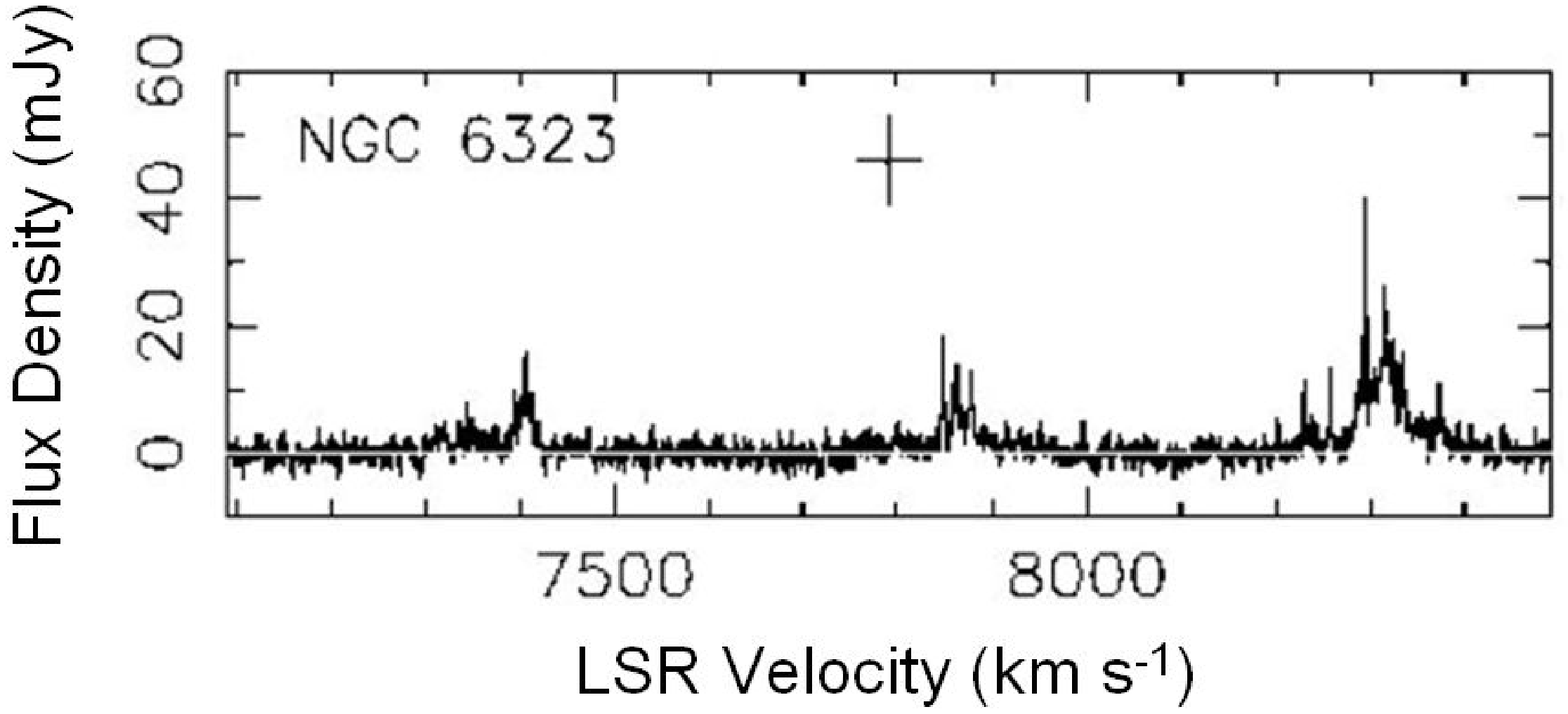}
  \caption{\textit{Left}: Composite spectrum of the velocity range over which $v=1, J=1 \to 0$ SiO maser emission was found with the VLA toward the Galactic center. That this is a ''mosaic'' obtained with seven different frequency settings was necessitated by the present VLA correlator's limited capabilities (from \cite{Reid_etal2007}). \textit{Right}: Spectrum of the 22.2 GHz \hzo\ line toward the active galaxy NGC 6323 taken with a wideband spectrometer at the Green Bank Telescope. The channel spacing is 0.33 \kms\  (from \cite{Braatz_etal2004}) To cover the 1100 \kms\ with emission at the same resolution with the current VLA correlator would require 37 settings.}\label{fig:spectra}
\end{center}
\end{figure}

%%\begin{figure}
%\includegraphics[height=2in,width=3in,angle=10]{fig02.eps}
%\includegraphics[width=8cm]{evlasensitivity.eps}
%  \caption{The solid lines give the continuum (left) and spectral line (right) sensitivities of the EVA %while the crosses mark the sensitivities of the current VLA  at the frequencies of its bands. While the huge %increases in continuum sensitivity, in particular at higher frequencies, are largely due to the EVLA's %increased bandwidth, the improvements in spectral line sensitivity are due to improved receiver %systems.}\label{fig:sensitivity}
%\end{figure}

\section{\label{vla}The Expanded Very Large Array}
Throughout its operation the VLA has made significant contributions to maser science, mainly through observations of ground-state hydroxyl (OH) and water vapor (\hzo) masers and, from the 1990s on, of SiO masers. %whose frequencies near 1.7 and 22.2 GHz fall into the traditional VLA (L and K) bands. Since the 1990s, the %Q band upgrade also allows observations of SiO masers at frequencies near 43 GHz.

%Let us emphasize that,
If phase calibration can be achieved, for maser surveys and time monitoring an interferometer is much preferred to a single dish telescope equipped with receivers of comparable quality. This is particularly true for the VLA, for the following reasons: (a) At K and Q band (\hzo\ and SiO masers), the VLA has a larger collecting area than any single dish (1.7 times that of a 100 m telescope). At longer wavelengths, the Arecibo 300m telescope has a much larger collecting area but can only observe a much smaller part of the sky. (b)  Atmospheric disturbances do not correlate. Consequently, spectra produced from interferometer data have higher quality.    (c) All of the VLA's receivers are available at all times. This is important for monitoring and not generally true for other observatories. (d) Importantly, even in its lowest resolution configuration (D-array) and, in particular at its maximum A-array resolutions around $1''$, $0\as08$, and $0\as04$ at 1.7, 22.2, and 43 GHz, respectively, the VLA yields good enough
\textit{relative} positions to get a detailed picture of the distribution of individual maser components and good enough \textit{absolute} positions for VLBI correlators.

Representing 1970's technology, the VLA, despite its continuing success, has many shortcomings, preventing it from being the ideal spectral-line and in particular maser survey instrument.
The Expanded Very Large Array (EVLA) will overcome many of the VLA's limitations. While using the existing VLA's infrastructure (antennas, railroad tracks, general logistics), %thanks to improvements enabled by the %progress in receiver technology and digital electronics
it will be equipped with more sensitive receivers with much larger bandwidth, full spectral coverage from 1 to 50 GHz, an immensely more powerful correlator and much faster electronics.

Comprehensive, detailed information on all technical aspects of the EVLA can be found on the EVLA Project Book WWW site: http://www.aoc.nrao.edu/evla/pbook.shtml
%In the following we compare %the most crucial improvements in more detail

%\subsection{\label{limitations}Limitations of the present VLA}
\subsection{\label{frequencycoverage}Frequency Coverage}
%As mentioned above, g
Given the bandwidths of receivers available in the 1970s, the VLA has a number of discrete wavelength bands each of limited width (see Fig. \ref{fig:evlacoverage}). Two of these, the 18--21 cm L band and 1.3 cm K band, contain important maser lines from the OH and the \hzo\ molecules, respectively. The 7 mm Q band added in the mid-1990s gave access to SiO maser lines. The very widespread 6.7 GHz (4.5 cm) \meth\ line, detected in 1991, and the 6.0 GHz (5 cm) OH hyperfine structure lines, excellent probes of the Zeeman effect, are not covered as aren't many other maser lines, e.g., from the \nhhh\ molecule (see Table 1).

% KMM stopped here

A key EVLA requirement is continuous frequency coverage from 1 to 50 GHz. This will be met with 8 bands (see Fig. \ref{fig:evlacoverage}): Two existing (K, Q), four replaced (L, C, X, U), and two new (S, A) bands. The  existing meter-wavelength bands, P (90 cm) and 4 (4 m) are retained with no changes. K and Q band will benefit from greatly improved receivers. In Fig. \ref{fig:evlacoverage} the dark stripes show the frequency coverage of the present VLA while the dark plus light stripes show that of the EVLA. All bands will have significantly wider instantaneous width than in the current VLA. In the highest frequency (K and Q) bands the bandwidth will be $2\times8$ GHz each (RCP+LCP); compare to today's $2\times100$ MHz! While affording fantastic continuum sensitivities, all the maser lines listed in Table 1 will be observable. In Fig. \ref{fig:evlacoverage} the frequencies of the 6.7 and 12.2 GHz \meth\ and the 6.0 GHz OH lines are marked. Also indicated is the EVLA's virtually limitless redshift coverage of the 22.2 GHz \hzo\ line. Note that the (by far) highest redshift \hzo\ line yet detected (from the $z=0.660$ QSO SDS J0804+3607) is shifted to 13.4 GHz (\cite{BarvainisAntonucci2005}).

\subsection{\label{vlacorrelator}The Correlator}
Most higher frequency (K and Q band) maser observations with the VLA have been severely limited by its 1970s technology correlator. The VLA correlator has a maximum bandwidth of 50 MHz per intermediate frequency (IF) band. In unsplit mode (i.e. 1 IF) it has a maximum of 16 frequency channels, which reduces to 8 per band if we chose a 2 IF mode, e.g., right plus left circular polarization and 4 per band for a $2\times$(RCP+LCP) = 4 IF mode, where two IF pairs with different frequencies (which have to be within 500 MHz of each other) are employed. For simplicity's sake let us only consider the 1 IF (1 polarization) mode for the 22.2 GHz \hzo\ line: To adequately resolve individual narrow maser features (and to avoid ``ringing'' caused by poorly resolved strong lines), we require a velocity resolution of 0.2 \kms, corresponding to 14.8 kHz, which is close to a correlator configuration that affords 256 channels spaced by 12.2 \kms. Considering that, due to band edge roll-off, only the inner 70\%\ of the bandpass are usable, the total velocity range that can be covered is 30 \kms, which is inadequate for many interstellar \hzo\ masers, which frequently have velocities spread over many tens of \kms, even if the systemic velocity of the source in question is known. Increasing the bandwidth by a factor of 2 results in half the channels and an inadequate four times worse resolution of 48.8 kHz (0.66 \kms) since we are dealing with a ``recirculating'' correlator. For SiO masers (Q band) the situation is even much worse, in particular for the kind of searches described in \S\ref{targeted}, in which the stars' velocities a are a priori unknown.

Extragalactic \hzo\ megamaser spectra frequently consist of narrow features with a width comparable to that of Galactic maser features ($\approx 1$ \kms; see Fig. 2). For the most interesting \hzo\ megamaser sources (those with NGC 4258-like Keplerian rotation signatures; see \cite{Miyoshi_etal1995}) maser features are found around the systemic velocity as well as at velocities offset by up to $\sim 1000$ \kms\ of the latter. %($\approx \pm 1000$ \kms in NGC 4258).
Consequently, when searching for new such sources one needs to scan a total range of $\approx 2000$ \kms, a tedious job with 30 \kms\ increments. The situation is aggravated by the fact that these masers are variable and many of their high velocity components are weak, which requires repeated and deep observations. Fig. \ref{fig:spectra} shows an \hzo\ spectrum taken toward the AGN at the center of NGC 6323 taken with the Green Bank telescope equipped with 200 MHz wide spectrometer with a channel spacing corresponding to 0.33 \kms.

The EVLA's correlator will overcome all of the described problems.
%Its The highest frequency bands (K and Q) will have bandwidths of $2\times8$ GHz (RCP + LCP) each. Choosing %the whole bandwidth the EVLA correlator affords 16384 channels or 8192 channels per IF. Recirculation, i.e., %trading numbers of channels against bandwidth provides higher spectral resolution. As an example, say, we %require a frequency spacing of 15.3 kHz (0.21 \kms\ at 22.2 GHz). This could achieved with 65536 channels %and a bandwidth of 1 GHz, resulting in a velocity coverage of 13483 \kms. This is (much) more than enough to %cover the velocity range of a galaxy. However, another application could be to search for multiple maser %lines (from Table 1) simultaneously with a bandwidth of a few hundred \kms\ each (see \S\ref{ultimate}).

\subsection{``Dead times''}
Every change in a VLA observation, be it a change in position or in the frequency setup, is afflicted by a ``dead time'' of 20 seconds, which derives mainly from limitations of the ancient electronics. This makes fast surveys in which integration times per position are small compared to these dead times extremely inefficient.  As an example, the NVSS was conducted as discrete snapshot pointings with each pointing taking 30 s of which only 7.7 s was actual dwell time (\cite{Condon_etal1998}).

On-the-fly (OTF) observations (see \S\ref{blindsurveys}) are impossible with the current VLA. A glimpse at the power of fast-scanning OTF interferometer surveys was recently provided by an 18 GHz  continuum survey with a single baseline of the Australia Telescope Compact Array (ATCA).  \cite{Ricci_etal2004} covered 1216 square degrees with an effective dwell time of 0.2 seconds, resulting in an rms noise of 15 mJy, they obtained 221 confirmed detections $> 60$ mJy.

For the EVLA very low overhead levels are a design feature.
\clearpage
\begin{table}[!t] %{lllc}
\footnotesize
 \begin{center}
 \caption{\label{maserlines}Astronomical Maser Transitions Observable with the EVLA}
 \begin{tabular}{llll}\hline
%\hline
Species                 & Transition & Frequency (MHz) & Number/Type \\
%\hline
OH~~$^2\Pi_{3/2}$       &$J={3\over2},$ $F=1\to2$&\d\d1612.2310(2)       &\textit{T}~SFR/\textit{H}~O-CSE   \\
                        &\phantom{$J={3\over2},$} $F=1\to1$     &\d\d1665.4018(1)       &\textit{H}~SFR/\textit{T}~O-CSE   \\
                        &\phantom{$J={3\over2},$} $F=2\to2$     &\d\d1667.3590(1)       &\textit{H}~SFR/\textit{T}~O-CSE   \\
                        &\phantom{$J={3\over2},$} $F=2\to1$     &\d\d1720.5300(1)       &\textit{H}~SFR            \\
                        &$J={5\over2},$ $F=2\to2$               &\d\d6030.747(5)        &\textit{T}~SFR            \\
                        &\phantom{$J={5\over2},$} $F=3\to3$     &\d\d6035.092(5)        &\textit{T}~SFR            \\
                        &$J={7\over2},$ $F=4\to4$               &\d13441.4173(2)        &\textit{S}~SFR/\textit{T}~SNR    \\
OH~~$^2\Pi_{1/2}$       &$J={1\over2},$ $F=0\to1$               &\d\d4660.242(3)  &\textit{T}~SFR    \\
                        &\phantom{$J={1\over2},$} $F=1\to1$     &\d\d4750.656(3)  &\textit{T}~SFR            \\  &\phantom{$J={1\over2},$} $F=1\to0$ &\d\d4765.562(3)  &\textit{T}~SFR            \\
%&\\
H$_2$O                  &\JK\d$6_{16}\to5_{23}$ &\d22235.08(2)             &\textit{H}~SFR/\textit{H}~O-CSE      \\
%&\\
CH$_3$OH$^{\rm a}$&\Jk\d$5_{1}\to6_{0}A^{+}$  &\d\d6668.5192(8)         &\textit{H}~SFR/Cl~II            \\
        &\pJk\d$9_{-1}\to8_{-2}E$   &\d\d9936.202(4)                    &\textit{1}~SFR/Cl~I  (W33-Met)         \\
        &\pJk\d$2_{0}\to3_{-1}E$    &\d12178.597(4)                     &\textit{H}~SFR/Cl~II            \\
        &\pJk\d$2_{1}\to3_{0}E$      &\d19967.3961(2)                   &\textit{S}~SFR/Cl~II            \\
        &\pJk\d$9_{2}\to10_{1}A^{+}$ &\d23121.0242(5)                    &\textit{S}~SFR/Cl~II            \\
%        &\pJk\d$3_{2}\to3_{1}E$      &\d24928.70(10)  &SFR/Cl~I            \\
%        &\pJk\d$4_{2}\to4_{1}E$      &\d24933.468(2)  &SFR/Cl~I            \\
%        &\pJk\d$5_{2}\to5_{1}E$      &\d24959.080(2)  &SFR/Cl~I            \\
        &\pJk\d$6_{2}\to6_{1}E$$^{\rm b}$ &\d25018.1225(4)   &\textit{T}~SFR/Cl~I            \\
        &\pJk\d$7_{2}\to7_{1}E$      &\d25124.8719(4)   &\textit{T}~SFR/Cl~I            \\
%        &\pJk\d$8_{2}\to8_{1}E$      &\d25294.411(3)   &SFR/Cl~I           \\
%        &\pJk\d$9_{2}\to9_{1}E$      &\d25541.43(10)   &SFR/Cl~I         \\
%        &\pJk$10_{2}\to10_{1}E$      &\d25878.18(10)   &SFR/Cl~I          \\
%        &\pJk$12_{2}\to12_{1}E$      &\d26847.27       &SFR/Cl~I          \\
%        &\pJk$13_{2}\to13_{1}E$      &\d27472.58      &SFR/Cl~I            \\
%        &\pJk$14_{2}\to14_{1}E$      &\d28169.52      &SFR/Cl~I            \\
%        &\pJk$15_{2}\to15_{1}E$      &\d28905.85       &SFR/Cl~I           \\
        &\pJk\d$8_{2}\to9_{1}A^{-}$  &\d28969.942(50) &\textit{S}~SFR/Cl~I            \\
%        &\pJk$16_{2}\to16_{1}E$      &\d29637.11       &SFR/Cl~I           \\
%        &\pJk$17_{2}\to17_{1}E$      &\d30308.08       &SFR/Cl~I           \\
        &\pJk\d$4_{-1}\to3_{0}E$     &\d36169.265(30)  &\textit{T}~SFR/Cl~I            \\
        &\pJk\d$7_{-2}\to8_{-1}E$    &\d37703.700(30)  &\textit{S}~SFR/Cl~II            \\
        &\pJk\d$6_{2}\to5_{3}A^{+}$  &\d38293.268(50)  &\textit{S}~SFR/Cl~II            \\
        &\pJk\d$6_{2}\to5_{3}A^{-}$  &\d38452.677(50)  &\textit{S}~SFR/Cl~II            \\
        &\pJk\d$7_{0}\to6_{1}A^{+}$  &\d44069.410(10)  &\textit{T}~SFR/Cl~I            \\
%&\\
NH$_3$$^{\rm c}$  &\JKN\d(9,6)                &\d18499.390(5)   &\textit{S}~SFR            \\
        &\pJKN\d(6,3)               &\d19757.538(5)   &\textit{S}~SFR            \\
%        &\pJKN\d(7,5)               &\d20804.830(5)   &SFR            \\
%        &\pJKN(10,8)                &\d20852.527(5)   &SFR            \\
%       &\pJKN(11,9)                &\d21070.739(5)   &SFR            \\
%        &\pJKN\d(5,4)               &\d22653.022(5)   &SFR            \\
%        &\pJKN\d(6,5)               &\d22732.429(5)     &SFR            \\
%        &\pJKN\d(9,8)               &\d23657.471(5)     &SFR            \\
        &\pJKN\d(3,3)               &\d23870.1296(1)         &\textit{S}~SFR            \\
%&\\
%$^{15}$NH$_3$&\JKN\d(4,3)           &\d21673.91             &SFR         \\
%        &\pJKN\d(3,3)               &\d22789.421(1)         &SFR         \\
%        &\pJKN\d(4,4)               &\d23046.0158(2)        &SFR            \\
%&\\
HC$_3$N &$J=~$$1\to0$               &\d\d9098.1152 (2)          &\textit{1}~SFR            \\
%&\\
H$_2$CO &\JK~$1_{10}\to1_{11}$      &\d\d4829.6600           &\textit{S}~SFR            \\
%&\\
CH~~$^2\Pi_{1/2}$       &$J={1\over2},$ $F=0\to1$               &\d\d3263.795(3)        &\textit{T}~SFR            \\
                        &\phantom{$J={1\over2},$} $F=1\to1$     &\d\d3335.481(2)        &\textit{T}~SFR            \\
                        &\phantom{$J={1\over2},$} $F=1\to0$     &\d\d3349.194(3)        &\textit{T}~SFR            \\
%&\\
SiO~~$v=0$      &$J=~$$1\to0$      &\d43423.858(10)        &\textit{1~SFR} (Orion-KL)/\textit{S}~O-CSE      \\
SiO~~$v=1$      &$J=~$$1\to0$      &\d43122.079(21)        &\textit{3~SFR}/\textit{H}~O-CSE      \\
SiO~~$v=2$      &$J=~$$1\to0$      &\d42820.582(23)        &\textit{3~SFR}/\textit{H}~O-CSE      \\
SiO~~$v=3$      &$J=~$$1\to0$      &\d42519.373(27)        &\textit{S}~O-CSE      \\
$^{29}$SiO~~$v=0$&$J=~$$1\to0$     &\d42879.916(10)        &\textit{1}~SFR (Orion-KL)/\textit{S}~O-CSE      \\
$^{30}$SiO~~$v=0$&$J=~$$1\to0$     &\d42373.359(10)        &\textit{1}~SFR (Orion-KL)/\textit{S}~O-CSE      \\
%&\\
SiS     &$J=$$1\to0$               &\d18154.880(2)         &\textit{1}~C-CSE (IRC+10216)      \\
\hline
  \end{tabular}
  \end{center}
\suppressfloats
\suppressfloats
\suppressfloats
%\textit{Specific Notes}:\\
\scriptsize
%\tiny
$^{\rm a}$A comprehensive compilation of \meth\ maser frequencies has been published by \cite{Mueller_etal2004}. $^{\rm b}$Emission and/or absorption from lines in the $J_2\to J_1E$ series of \meth\ has been detected from $J = 3, \nu = 24928.70$ MHz to $J= 17, \nu = 30308.08$ MHz. Clear maser emission has been detected for the $J = 5$ to $J=9$ lines. The $J=6$ and 7 lines listed here are usually the strongest maser lines of this series. $^{\rm c}$Weak maser emission from 6 other non-metastable ($J\neq K$) $^{14}$NH$_3$ and 3 $^{15}$NH$_3$ lines between 20.8 and 23.7 GHz has been reported in the literature.\\
\textit{General remarks:}
Except for the high excitation \hzo, \nhhh, and vibrationally excited SiO lines, many lines may also be observable in (quasi)thermal emission and/or absorption in regions that are not conducive to maser excitation. See  F. Lovas' WWW site for references to the laboratory measurements of the frequencies and for the first astronomical measurements of the lines (http://physics.nist.gov/cgi-bin/micro/table5/start.pl).
%For that information on methanol maser lines, see \cite{Mueller_etal2005}.
The numbers in parentheses following the frequencies denote the measurement error in the
last significant digit(s).
%Note that frequencies given, e.g., in compilations
%such as the JPL line catalog
%catalog (http://spec.jpl.nasa.gov/),
%are generally determined from
%Hamiltonian fitting and generally less accurate then the frequencies given
%here, which are mostly from  laboratory measurements.
Abbreviations for different maser environments are SFR for star-forming
regions and O-CSE and C-CSE for oxygen-rich and carbon-rich circumstellar envelopes, respectively. Maser emission from the 1720 MHz OH line has also been found in supernova remnants (SNR). A leading ``\textit{H}'' means that there are hundreds of sources known, a ``\textit{T}'' means tens ($<100$) and an ``\textit{S}'' means some ($<10$). A ``\textit{1}'' means that only a single source has been found to show maser action and that source is listed. Apart from Orion-KL, SiO masers have been found in the \textit{3} SFRs W51N and Sgr B2(M). Methanol masers are marked as Class (Cl) I or to II; see
\cite{Menten1991}.
\end{table}

\subsubsection{\label{ultimate}The power of the EVLA: The ultimate Galactic plane star formation survey}
To demonstrate the awesome power of the EVLA let us consider a blind survey to find and characterize star-forming regions in the Galaxy. Let us first consider the 4--8 GHz EVLA C band. This band contains the frequency (6.7 GHz) of the by far most prominent class II \meth\ maser line, the presence of which is a unique tracer of high-mass star formation. The power of sensitive \meth\ maser surveys in this line has recently been demonstrated by a blind survey with the Arecibo 300m telescope: \cite{Pandian_etal2007} (see also these proceedings) observed a $b = \pm0\decdeg41$ wide strip from $l = 35\decdeg2$ to $53\decdeg7$ and found, at a completeness level of 0.27 Jy, a total of 86 sources, 48 of them new detections. A large scale single-dish survey covering all of the Galactic plane with an 8 channel array receiver on the Parkes and the Jodrell Bank telescopes is currently under way (\cite{Cohen_etal2007}).

A possible fully sampled C band Galactic plane survey would not only cover the 6.7 GHz methanol line but, in addition, all four hyperfine structure (hfs) lines of rotationally excited OH near 6.0 GHz and the three OH hfs lines near 4.7 GHz (see Table \ref{maserlines}).
The two strongest of the former, at 6030 and 6035 MHz, are excellent lines to study Zeeman splitting and determining magnetic field strengths in ultracompact HII regions (UCHIIRs) where they frequently exhibit maser emission. In more developed, compact HII regions they are seen in absorption (see, e.g., \cite{Fish_etal2006}).

Most interestingly, one would also be able to detect continuum emission from extremely weak UCHIIRs.
%Splitting the 4 GHz bandwidth in four 1 GHz chunks one would reach (in 10 sec) an rms noise of 83 %$\mu$Jy~beam$^{-1}$.
With the 4 GHz bandwidth one would reach in 10 sec an rms noise of 41 $\mu$Jy~beam$^{-1}$.
Note that the famous dual frequency \cite{WoodChurchwell1989} targeted survey of 75 UCHIIRs reached a comparable noise level in 4 minutes!
%typical theoretical rms value of 0.09 \mjyb at 4.86 GHz \textit{in 4 minutes} of observing time (and 0.28 %\mjyb\ at 14.94 GHz). The  real sensitivity of the latter, also being a snapshot survey, was determined by dynamic range limitations and roughly 3 times higher than the theoretical values and a similar factor should be allowed for in the EVLA survey discussed here.

Splitting the continuum band in, e.g., four parts one would be able to determine the spectral properties of the continuum emission between 4 and 8 GHz, a range in which compact and many ultracompact HII region ''turn over'' from an optically thick spectrum with flux density, $S$, $\propto \nu^2$, to an optically thin spectrum ($S \propto \nu^{-0.1}$). The dynamics of the ionized gas could be studied from the velocities of recombination lines.
Note also, that with 50\%\ fractional
bandwidth coverage, the $uv$-plane would be heavily filled, which should result in improved imaging of the continuum emission (``bandwidth synthesis'').

Moreover, it would be possible to observe formaldehyde (\hhco) absorption in the ubiquitous 4.8 GHz ground-state doublet line against the UCHIIRs' continuum emission and, thus, resolve the kinematic distance ambiguity for the detected star-forming regions. To emphasize, by setting spectral windows around the lines in question in will be possible to image all of the mentioned lines (and more) \textit{and} the continuum \textit{simultaneously}.

The ideal complement to this C band survey would be a K band (18--26 GHz) survey which would detect very low luminosity \hzo\ masers throughout the galaxy. Note that in contrast to \meth\ and OH masers, \hzo\ masers are found in high- \textit{and} low-mass star-forming regions. Known masers in the nearby $\rho$ Oph or Perseus low-mass stars forming regions at distances, $D$, of 160 and 350 pc, respectively, have been found with flux densities of hundreds or even thousands of Jy (\cite{Claussen_etal1996}). Such masers placed at the distance of the Galactic center ($D = 8$ kpc) would have flux densities of a few tenths to a few Jy and be detectable in our survey. Such water masers would, in fact, be the only means to find evidence for active low-mass star formation at distances greater than a few kpc. \footnote{We note however, that the luminosity function of \hzo\ masers is poorly constrained at the lowest luminosity end (\cite{Palla_etal1993}). Therefore such detections would deliver little information for putting statistical or other constraints on low-mass star formation.}

Apart from detecting \hzo\ masers, the K band survey would highly efficiently detect hypercompact HII regions (HCHIIRs; see \cite{Kurtz2005}). These represent highly compact ($<50$ AU) radio continuum emission from embedded protostellar objects with early B to O star luminosities that have an optically thick spectrum up to the highest radio frequencies. With typical K band flux densities of a few mJy or even less (at a few kpc distance) the radio emission is thought to be ``choked off'' by accretion in these sources (\cite{Walmsley1995}) and its intensity cannot be predicted from their total luminosity. Representing the earliest (pre-UCHIIR) stages of high-mass star formation, observations of HCHIIRs are of central interest to constrain high-mass protostellar evolution schemes.

Also covered by the K-band survey would be simultaneous searches for all of the maser lines in that range listed in Table \ref{maserlines} and sensitive searches for absorption lines from  ammonia and other molecules toward the continuum emission of stronger UCHIIRs.

An important by-product of the K band survey would be a list of high frequency calibrator candidates whose suitability, i.e., compactness, could be tested with multi-wavelength data, more extended configuration EVLA  and/or VLBI observations. A dense net of calibrator sources will be highly useful for all high frequency EVLA observations and mandatory for ALMA, for which fast-switching (requiring a nearby calibrator) will be a standard observing procedure and not an exception as currently with the VLA.

If one conducted the C band survey in the B-configuration and the K band survey in the three times more compact C-configuration, both data sets could be restored with comparable synthesized beams around $1''$, which would be highly desirable for meaningful determinations of  band-to-band spectral index distributions.
%At such resolution it would also be possible to study the clustering properties of the masers and their relation to other phenomena, in particular to possible pumping sources.

\subsubsection{What the EVLA will not do}
The EVLA will not provide any enhancements over the VLA for frequencies below 1 GHz. This means, for example, that searches for high-redshift 1667 MHz OH megamasers will be restricted to the frequency range of the current P band system (306--340 MHz), corresponding to $z = $ 3.90--4.45), which will be retained unchanged. While a VLA search for OH gigamaser emission in the $z = 3.911$ broad absorption line quasar APM08279+5255, a well-known source of redshifted dust and CO emission, was unsuccessful (\cite{Ivison2006}), we note that this was the \textit{only} known such source to fall into this redshift band. Given this situation, the detection of high-$z$ OH gigamasers at $z > 0.67$ may have to await the Square Kilometer Array (see \S\ref{ska}).

Moreover, very unfortunately, EVLA Phase II, consisting of physically extending the VLA by adding additional antennas to bridge the gap from VLA to VLBA resolution (''The New Mexico Array'') was not funded.
For a long time MERLIN, will remain the only option to cover that interesting regime at radio wavelengths.
 MERLIN is currently undergoing an upgrade, resulting, a.o., in wider bandwidth receivers, that will enhance its sensitivity up to 30 fold (``e-MERLIN'', \cite{Garrington_etal2004}).

Finally, we emphasize the importance of the ATCA, the only versatile operating radio wavelength interferometer in the southern hemisphere, which will greatly increase with the advent of ALMA. The successful recent 7 mm upgrade and the soon too be installed much wider band correlator are very significant milestones.
\suppressfloats
\suppressfloats
\suppressfloats

\section{The Atacama Large Millimeter Array}
ALMA in its highest resolution configurations at 345 GHz, a wavelength at which weather conditions should allow observations for most of the time, in current planning will have a synthesized beam width of 12 milliarcseconds (mas),  $\approx 3$ times better than the highest VLA resolution. For  maser science with ALMA, see A. Wootten's contribution to these proceedings.  A few additional thoughts follow. %on (sub)millimeter  \meth, SiO, \hzo, and HCN masers follow.
\suppressfloats
\suppressfloats
\suppressfloats

\subsection{Methanol masers}
 So far, a total of 19 class I and class II \meth\ maser lines with frequencies between 84 and 230 GHz have been discovered in star-forming regions (\cite{Mueller_etal2004}). Class II \meth\ masers show mas structure, i.e., a few AU at a few kpc and are VLBI targets (see, e.g., the contributions by Reid et al. and Xu et al. to these proceedings). In contrast, class I \meth\ masers are heavily resolved at the highest VLA resolutions: 25 GHz maser spots in the nearby ($\approx 400$ pc) Orion region have sizes between 20 and 60 AU, or 50 -- 150 mas (\cite{Johnston_etal1997}) and VLBI observations of class I masers have not produced fringes (\cite{Lonsdale_etal1998}). With resolutions intermediate between VLA and VLBI values, ALMA observations of class I masers will reveal their morphology. Since existing observations suggest that they mark interface regions between protostellar outflows and dense ambient gas (\cite{Menten1996}) such observations will be very interesting.

 \subsection{Water masers}
 Apart from the most widespread 22.2 GHz $6_{16}\to5_{23}$  transition, a number
 of (sub)millimeter lines have been discovered from the vibrational ground state as well as from the lowest vibrationally excited state (see the contribution of Hunter et al. to these proceedings). Excitation calculations predict temperature and density ranges in which one might expect inversion of specific lines (\cite{NeufeldMelnick1991}).
 %In turn, the simultaneous occurrence of maser action in certain lines should help constraining physical %conditions.
 To exploit this, one has to ascertain that maser emission from different lines indeed arises from identical regions. Up to now this was not possible as virtually all submillimeter \hzo\ data came from single dish observations. Now, (quasi)simultaneous interferometric observations of the 22.2 and 321 GHz \hzo\ maser lines reported in these proceedings by Patel et al. indeed point to a more complicated picture. They find that in the Cepheus A star-forming region the \hzo\ maser emission from the much higher excitation 321 GHz line (1846 K above the ground-state) comes from different regions than the emission in the lower excitation 22.2 GHz (643 K) line. Needless to say, this is a wide field awaiting ALMA.

\subsection{HCN masers}
Maser emission from hydrogen cyanide (HCN) has been detected toward several mass-losing carbon stars.
%While in some objects, the $J=1 - 0$ transition in the vibrational ground-state shows signs of maser %emission,
Most of the maser lines are emitted from the vibrationally excited $\nu_2$ state.
For all circumstellar (SiO, \hzo, and HCN) masers it should be noted that highest resolution ALMA observations will not only reveal the structure of the maser distributions but will also resolve the stellar submillimeter photospheres of their host AGB stars (Miras) out to several hundred pc and even of further away red supergiants.

\section{\label{ska}The Square Kilometer Array}
See A. Green's contribution to these proceedings for a description of maser science with the SKA. Again, a  few more thoughts:

The usefulness of OH mega- (or rather giga-)masers as a probe of violent star formation at early epochs, complementary to (sub)millimeter dust and molecular and radio continuum studies, has been discussed in a number of studies (\cite{Briggs_etal1998, Townsend_etal2001, Ivison2006}). The SKA will have the potential to detect OH gigamaser emission at all redshifts of potential interest.

The SKA will open new vistas for maser astrometry  with microarcsecond accuracies (\cite{FomalontReid2004}) and the collecting area necessary to image NGC 4248-like systems into the Hubble flow, allowing a high-accuracy determination the Hubble constant (see  L. Greenhill's contribution to these proceedings).
%\section{Conclusions}\label{sec:concl}

\begin{acknowledgments}
I'm very grateful to Mark Reid for a thorough reading of the manuscript of this paper and to him and Rick Perley for useful suggestions.
\end{acknowledgments}
\suppressfloats
\suppressfloats
\suppressfloats


\begin{thebibliography}{}

\bibitem[Barvainis \& Antonucci (2005)]{BarvainisAntonucci2005} Barvainis, R.,
\& Antonucci, R.\ 2005, \textit{ApJ}, 628, L89

\bibitem[Benjamin et al. (2003)]{Benjamin_etal2003} Benjamin, R.~A., et
al.\ 2003, \textit{PASP}, 115, 953

\bibitem[Braatz et al. (2004)]{Braatz_etal2004} Braatz, J.~A., Henkel,
C., Greenhill, L.~J., Moran, J.~M., \& Wilson, A.~S.\ 2004, \textit{ApJ}, 617, L29

\bibitem[Briggs (1998)]{Briggs_etal1998} Briggs, F.~H.\ 1998, \textit{A\&A}, 336,
815

\bibitem[Caswell \& Haynes (1987)]{Caswell_etal1987} Caswell, J.~L., \&
Haynes, R.~F.\ 1987, \textit{Austr. J. of Physics}, 40, 215


%\bibitem[Caswell \& Haynes (1983)]{CaswellHaynes1983} Caswell, J.~L., \&
%Haynes, R.~F.\ 1983, \textit{Austr. J. of Physics}, 36, 361


%\bibitem[Caswell et al. (1980)]{Caswell_etal1980} Caswell, J.~L., Haynes,
%.~F., \& Goss, W.~M.\ 1980, \textit{Austr. J. of Physics}, 33, 639

\bibitem[Caswell et al. (1989)]{Caswell_etal1989} Caswell, J.~L.,
Batchelor, R.~A., Forster, J.~R., \& Wellington, K.~J.\ 1989, \textit{Austr.
J. of Physics}, 42, 331


%\bibitem[Caswell et al. (1983)]{Caswell_etal1983b} Caswell, J.~L.,
%Batchelor, R.~A., Forster, J.~R., \& Wellington, K.~J.\ 1983, \textit{Austr.
%J. of Physics}, 36, 443


%\bibitem[Caswell et al. (1983)]{Caswell_etal1983a} Caswell, J.~L.,
%Batchelor, R.~A., Forster, J.~R., \& Wellington, K.~J.\ 1983, \textit{Austr.
%J. of Physics}, 36, 401


\bibitem[Caswell et al. (1995)]{Caswell_etal1995} Caswell, J.~L., Vaile,
R.~A., Ellingsen, S.~P., Whiteoak, J.~B., \& Norris, R.~P.\ 1995, \textit{ MNRAS},
272, 96

\bibitem[Churchwell et al. (1977)]{Churchwell_etal1977} Churchwell, E.,
Witzel, A., Huchtmeier, W., Pauliny-Toth, I., Roland, J., \& Sieber, W.\
1977, \textit{A\&A}, 54, 969

\bibitem[Claussen et al. (1996)]{Claussen_etal1996} Claussen, M.~J.,
Wilking, B.~A., Benson, P.~J., Wootten, A., Myers, P.~C., \& Terebey, S.\
1996, \textit{ApJS}, 106, 111

\bibitem[Cohen et al. (2007)]{Cohen_etal2007} Cohen, R.~J., et al.\
2007, IAU Symposium, 237, 403

\bibitem[Condon et al. (1998)]{Condon_etal1998} Condon, J.~J., et al.\ 1998, \textit{AN}, 115, 1693

\bibitem[Deacon et al. (2007)]{Deacon_etal2007} Deacon, R.~M., Chapman,
J.~M., Green, A.~J., \& Sevenster, M.~N.\ 2007, \textit{ApJ}, 658, 1096

\bibitem[Fish et al. (2006)]{Fish_etal2006} Fish, V.~L., Reid, M.~J.,
Menten, K.~M., \& Pillai, T.\ 2006, \textit{A\&A}, 458, 485

\bibitem[Fomalont \& Reid (2004)]{FomalontReid2004} Fomalont, E., \&
Reid, M.\ 2004, \textit{New Astronomy Review}, 48, 1473

\bibitem[Garrington et al. (2004)]{Garrington_etal2004} Garrington, S.~T.,
et al.\ 2004, \textit{Proc. SPIE}, 5489, 332

\bibitem[Habing et al. (2006)]{Habing_etal2006} Habing, H.~J.,
Sevenster, M.~N., Messineo, M., van de Ven, G., \& Kuijken, K.\ 2006, \textit{A\&A},
458, 151

\bibitem[Ivison (2006)]{Ivison2006} Ivison, R.~J.\ 2006, \textit{MNRAS},
370, 495

\bibitem[Johnston et al. (1997)]{Johnston_etal1997} Johnston, K.~J., et al.\
%Gaume, R.~A., Wilson, T.~L., Nguyen, H.~A., \& Nedoluha, G.~E.\
1997, \textit{ApJ}, 490, 758

\bibitem[Kurtz (2005)]{Kurtz2005} Kurtz, S.\ 2005, Massive Star
Birth: A Crossroads of Astrophysics, 227, 111

\bibitem[Lo (2005)]{Lo2005} Lo, K.~Y.\ 2005, \textit{ARA\&A}, 43, 625

\bibitem[Lonsdale et al. (1998)]{Lonsdale_etal1998} Lonsdale, C.~J., et
al.\ 1998, \textit{Bulletin of the American Astronomical Society}, 30, 1355

\bibitem[Menten (1991)]{Menten1991} Menten, K.~M.\ 1991, \textit{ApJ},
380, L75


%\bibitem[Menten (1991)]{Menten1991b} Menten, K.\ 1991, in \textit{Atoms, Ions
%and Molecules: New Results in Spectral Line Astrophysics}, 16, 119

\bibitem[Menten (1996)]{Menten1996} Menten, K.~M.\ 1996, Molecules
in Astrophysics: Probes \& Processes, 178, 163

\bibitem[Messineo et al. (2002)]{Messineo_etal2002} Messineo, M., Habing,
H.~J., Sjouwerman, L.~O., Omont, A., \& Menten, K.~M.\ 2002, \textit{A\&A}, 393, 115

\bibitem[Miyoshi et al. (1995)]{Miyoshi_etal1995} Miyoshi, M., et al.\ 1995,
\textit{Nature}, 373, 127

\bibitem[M{\"u}ller et al. (2004)]{Mueller_etal2004} M{\"u}ller,
H.~S.~P., Menten, K.~M., \& M{\"a}der, H.\ 2004, \textit{A\&A}, 428, 1019

\bibitem[Murakami et al.(2007)]{Murakami_etal2007} Murakami, H., et al.\
2007, ArXiv e-prints, 708, arXiv:0708.1796

\bibitem[Neufeld \& Melnick (1991)]{NeufeldMelnick1991} Neufeld, D.~A., \&
Melnick, G.~J.\ 1991, \textit{ApJ}, 368, 215

\bibitem[Norris et al. (1987)]{Norris_etal1987} Norris, R.~P., Caswell,
J.~L., Gardner, F.~F., \& Wellington, K.~J.\ 1987, \textit{ApJ}, 321, L159

\bibitem[Omont et al. (2003)]{Omont_etal2003} Omont, A., et al.\ 2003,
\textit{A\&A}, 403, 975

\bibitem[Palla et al. (1993)]{Palla_etal1993} Palla, F., Cesaroni, R.,
Brand, J., Caselli, P., Comoretto, G., \& Felli, M.\ 1993, \textit{A\&A}, 280, 599

\bibitem[Pandian et al. (2007)]{Pandian_etal2007} Pandian, J.~D.,
Goldsmith, P.~F., \& Deshpande, \textit{A.~A.}\ 2007, ApJ, 656, 255



%\bibitem[Price et al. (1995)]{Price1995}
%Price, S. D., \textit{Space Sci. Rev.} 1995, 74, 81

\bibitem[Reid et al. (2007)]{Reid_etal2007} Reid, M.~J., Menten,
K.~M., Trippe, S., Ott, T., \& Genzel, R.\ 2007, \textit{ApJ}, 659, 378

\bibitem[Ricci et al.(2004)]{Ricci_etal2004} Ricci, R., et al.\ 2004,
\textit{}, 354, 305

\bibitem[Schuller et al. (2003)]{Schuller_etal2003} Schuller, F., et al.\
2003, \textit{A\&A}, 403, 955


%\bibitem[Sevenster et al. (1997)]{Sevenster_etal1997a} Sevenster, M.~N.,
%Chapman, J.~M., Habing, H.~J., Killeen, N.~E.~B., \& Lindqvist, M.\ 1997,
%\textit{A\&AS}, 122, 79


%\bibitem[Sevenster et al. (1997)]{Sevenster_etal1997b} Sevenster, M.~N.,
%Chapman, J.~M., Habing, H.~J., Killeen, N.~E.~B., \& Lindqvist, M.\ 1997,
%\textit{A\&AS}, 124, 509

\bibitem[Sevenster et al. (2001)]{Sevenster_etal2001} Sevenster, M.~N., van
Langevelde, H.~J., Moody, R.~A., Chapman, J.~M., Habing, H.~J., \& Killeen,
N.~E.~B.\ 2001, \textit{A\&A}, 366, 481

%\bibitem[Te Lintel Hekkert et al. (1991)]{TeLintelHekkert_etal1991} Te Lintel
%Hekkert, P., Caswell, J.~L., Habing, H.~J., Haynes, R.~F., Haynes, R.~F.,
%\& Norris, R.~P.\ 1991, \textit{A\&AS}, 90, 327

\bibitem[Townsend et al. (2001)]{Townsend_etal2001} Townsend, R.~H.~D.,
Ivison, R.~J., Smail, I., Blain, A.~W., \& Frayer, D.~T.\ 2001, \textit{MNRAS},
328, L17

\bibitem[Walmsley (1995)]{Walmsley1995} Walmsley, M.\ 1995, Revista
Mexicana de Astronomia y Astrofisica Conference Series, 1, 137

\bibitem[Wood \& Churchwell (1989)]{WoodChurchwell1989} Wood, D.~O.~S., \&
Churchwell, E.\ 1989, \textit{ApJS}, 69, 831

%\bibitem[Zijlstra et al. (2001)]{Zijlstra_etal2001} Zijlstra, A.~A.,
%Chapman, J.~M., te Lintel Hekkert, P., Likkel, L., Comeron, F., Norris,
%R.~P., Molster, F.~J., \& Cohen, R.~J.\ 2001, \textit{MNRAS}, 322, 280
\bibitem[Zijlstra et al. (2001)]{Zijlstra_etal2001} Zijlstra, A.~A., et al.,\
%Chapman, J.~M., te Lintel Hekkert, P., Likkel, L., Comeron, F., Norris,
%R.~P., Molster, F.~J., \& Cohen, R.~J.\
2001, \textit{MNRAS}, 322, 280

\end{thebibliography}
\end{document}